\documentclass[aps,prl,12pt,showkeys]{revtex4}

\usepackage{graphicx}

\begin{document}
\title{\bf Heisenberg's wave packet reconsidered}
\author{J. Orlin Grabbe}
\email[Email: ]{quantum@orlingrabbe.com}
\date{September 11, 2005}

\begin{abstract}
This note shows that Heisenberg's choice for a wave function in his original paper on the uncertainty
principle is simply a renormalized characteristic function of a stable distribution with
certain restrictions on the parameters.  Relaxing Heisenberg's restrictions leads to a more general
formulation of the uncertainty principle. This reformulation shows quantum uncertainty can exist at a
macroscopic level. These modifications also give rise to a new form of Schr\"{o}dinger's wave equation
as the equation of a vibrating string. Although a heat equation version can also be given, the latter shows the
traditional formulation of Schr\"{o}dinger's equation involves a hidden Cauchy amplitude assumption.

\end{abstract}


\maketitle

\subsection{A generalized wave packet}

We begin by showing that Heisenberg's choice for a wave function in his original paper \cite{WH} on the uncertainty
principle is simply a renormalized characteristic function of a stable distribution,
$S_{\alpha,\beta}(x;m,c)$ with $\alpha = 2$ and $\beta = 0$, and location and scale parameters $m$ and
$c$.  Relaxing the assumptions on $\alpha, \beta$ so that $0<\alpha\le 2, \beta \ne 0$, leads to a
more general formulation of the uncertainty principle.  These modifications also give rise to a new form
of Schr\"{o}dinger's partial differential equation.

Consider the following wave packet $\psi(x,t)$, where at time $t = 0$, $\psi(x,0)$ has the form
\begin{equation}
\psi(x,0) = A_o \mbox{ exp}[imx-c|x|^{\alpha}],
\end{equation}
where
\begin{equation}
A_o = [\frac{\alpha (2c)^{\frac{1}{\alpha}}}{2\Gamma(\frac{1}{\alpha})}]^{\frac{1}{2}} .
\end{equation}
It is easy to see that $\psi(x,0)$ is normalized to unity:
\begin{equation}
\int_{-\infty}^{\infty} \psi*(x,0)\psi(x,0) dx = A_o^2 \int_{-\infty}^{\infty} \mbox{ exp}[-2c|x|^{\alpha}] dx 
= 2A_o^2 \int_0^{\infty} \mbox{ exp}[-2cx^{\alpha}] dx .
\end{equation}
Using the relation
\begin{equation}
\int_0^{\infty} y^k e^{-y^{\alpha}} dy = \frac{1}{\alpha}\Gamma(\frac{k+1}{\alpha})
\end{equation}
and making the substitution $u = (2c)^{\frac{1}{\alpha}}x$, we obtain
\begin{equation}
2 A_o^2 \int_0^{\infty} \mbox{ exp}[-2cx^{\alpha}] dx = \frac{2A_o^2}{(2c)^{\frac{1}{\alpha}}} \int_0^{\infty} e^{-u^{\alpha}} du = \frac{2A_o^2}{(2c)^{\frac{1}{\alpha}}} \frac{1}{\alpha} \Gamma(\frac{1}{\alpha}) = 1 .
\end{equation}
Now, the form of the wave packet in Eq.(1) can be compared to Heisenberg's original wave packet, denoted
here $H(x,0)$:
\begin{equation}
H(x,0) = (2\tau)^{\frac{1}{4}} \mbox{ exp}[2\pi i\sigma_o x - \pi \tau x^2] .
\end{equation}
If we make the substitutions
\begin{eqnarray}
2\pi \sigma_o = m\\
\pi \tau = c\\
\alpha = 2
\end{eqnarray}
in $\psi(x,0)$, the wave packet of Eq.(1), we obtain $H(x,0)$.  (Note that with $\alpha = 2$,
$A_o = [\frac{2 (2c)^{\frac{1}{2}}}{2\Gamma(\frac{1}{2})}]^{\frac{1}{2}} = (\frac{2c}{\pi})^{\frac{1}{4}}
= (2\tau)^{\frac{1}{4}}$ .)

Now let's derive the amplitude function of $\psi(x,0)$, which will necessarily also give us the amplitude
function of $H(x,0)$. First note that the log characteristic function of a stable distribution is 
\begin{eqnarray}
\mbox{ log }\varphi (z) = \mbox{ log }\int_{-\infty}^{\infty} \mbox{ exp}[ixz] dF(\frac{x-m}{c'}) \\
= imz - |c'|^{\alpha}|z|^{\alpha}[1 + i\beta(z/|z|) \mbox{tan}(\pi \alpha /2)], \mbox{ if } \alpha \ne 1\\
= imz - |c'|^{\alpha}|z|^{\alpha}[1 + i\beta(z/|z|) (2/ \pi) \mbox{log}|z|], \mbox{ if } \alpha = 1
\end{eqnarray}
where $m$ is a real number, $c' \ge 0$, $0<\alpha \le 2$, $|\beta| \le 1$. Proof of this theorem, due to
Khintchine and L\'{e}vy in 1936, may be found in \cite{PL} or \cite{GK}.  Here $\alpha$, the \emph{characteristic
exponent}, is essentially an index of peakedneess ($\alpha = 2$ for the normal or Gaussian distribution, $\alpha = 1$
for the Cauchy distribution).  The parameter $\beta$ is an index of skewedness ($\beta = 0$ for symmetric distributions).  The parameter $c'= c^{\frac{1}{\alpha}}$ is a scale parameter (the standard deviation when $\alpha = 2$).  Finally, $m$ is a location parameter (the mean if $\alpha > 1$; it is also the median or modal value of the distribution if $\beta = 0$).

For $\beta = 0$ we obtain the characteristic function of a \emph{symmetric} stable distribution, which
is identical to Eq.(1), if we omit the normalizing constant $A_o$.
Therefore, for the amplitude function of our wave packet, we take the Fourier transform, $A(z)$, of Eq.(1) to obtain
\begin{equation}
A(z) = \int_{-\infty}^{\infty} \psi(x,0) \mbox{ exp}[-ixz] dx = A_o \int_{-\infty}^{\infty} \mbox{ exp}[imx-c|x|^{\alpha}] \mbox{ exp}[-ixz] dx = A_o s_{\alpha,0}(z;m,c).
\end{equation}
In other words, we obtain a symmetric stable density function $s_{\alpha,0}(z;m,c) = dS_{\alpha,0}(z;m,c)$ with the
normalization constant $A_o$ for the amplitude function.  The symmetric stable density has $0<\alpha \le 2$, $\beta = 0$, and location and scale paramters $m$ and $c$, respectively.  Note that the amplitude function is normalized so that the integral of its \emph{square} is equal to 1. This involves the square of the probability density function $s_{\alpha,0}(z;m,c)$.

For Heisenberg's case where $\alpha = 2$, we may explicity solve for $A(z) = A(\sigma)$, which will be necessarily the
Gaussian density multiplied by a normalizing constant.  We reintroduce a factor of $2\pi$ to obtain
\begin{equation}
A(\sigma) = \int_{-\infty}^{\infty}(2\tau)^{\frac{1}{4}} \mbox{ exp}[2\pi i\sigma_o x - \pi \tau x^2] \mbox{ exp}[-2\pi ix\sigma] dx = (\frac{2}{\tau})^{\frac{1}{4}}  \mbox{ exp}[-\frac{\pi(\sigma-\sigma_o)^2}{\tau}] .
\end{equation}
This is Heisenberg's amplitude function.  That the integral of its square is 1 follows from:
\begin{equation}
\int_{-\infty}^{\infty} [A(\sigma)]^2 d\sigma = (\frac{2}{\tau})^{\frac{1}{2}} \int_{-\infty}^{\infty} \mbox{ exp}[-\frac{2\pi(\sigma-\sigma_o)^2}{\tau}] d\sigma = (\frac{2}{\tau})^{\frac{1}{2}} (\frac{\tau}{2\pi})^{\frac{1}{2}} 2 \int_{0}^{\infty} e^{-u^2} du = \frac{\sqrt \pi}{\sqrt \pi} = 1,
\end{equation}
where we have used the substitution $u = \sqrt{\frac{2\pi}{\tau}} (\sigma-\sigma_o)$ . Note that the usual normalizing
constant $\frac{1}{d\sqrt{2\pi}}$ for the Gaussian distribution (where $d$ is the standard deviation) has been absorbed into $A_o$.  So above and below, when we write the stable density $s_{\alpha,\beta}(z;m,c)$, we will understand the omission of the usual normalizing constant, and will consider only the normalizing $A_o$ in the product $A_o s_{\alpha,\beta}(z;m,c)$.  This will ensure that the square of the amplitude function is a probability
distribution.

\subsection{Alternative amplitude functions}

For $\alpha = 1$, which corresponds to the Cauchy distribution, the normalizing constant $A_o = c^{\frac{1}{2}} =
(c')^{\frac{\alpha}{2}} = (c')^{\frac{1}{2}}$, so the amplitude function is
\begin{equation}
A(z) = A_o s_{1,0}(z;m,c) = c^{\frac{1}{2}} \sqrt{\frac{2}{\pi}} \frac{c}{c^2 + (z-m)^2} ,
\end{equation}
where we have removed a division by $\sqrt{2\pi}$ in the usual statement of the Cauchy.  That this is the
correct normalization for the amplitude function in Eq.(16) follows from the integral:
\begin{equation}
\int_{-\infty}^{\infty} [A(z]^2 dz = \frac{2}{\pi} \int_{-\infty}^{\infty} \frac{c^3}{[c^2 + (z-m)^2]^2} dz =
\frac{2}{\pi} \int_{-\infty}^{\infty} \frac{1}{[1 + y^2]^2} dy ,
\end{equation} 
where we have used the substitution $y = \frac{z-m}{c}$.  We may now appeal to the relations, for $a,c>0$ and $n$
a positive integer:
\begin{equation}
\int \frac{dx}{(ax^2+c)^n} = \frac{1}{2(n-1)c} \frac{x}{(ax^2+c)^{n-1}} + \frac{2n-3}{2(n-1)c} \int \frac{dx}{(ax^2+c)^{n-1}}
\end{equation}
and
\begin{equation}
\int \frac{dx}{ax^2+c} = \frac{1}{\sqrt{ac}} \mbox{ tan}^{-1} [x \sqrt{\frac{a}{c}}] .
\end{equation}
Thus we get
\begin{equation}
\frac{2}{\pi} \int_{-\infty}^{\infty} \frac{dy}{[1 + y^2]^2} = \frac{2}{\pi} \frac{1}{2} \int_{-\infty}^{\infty} \frac{dy}{1+y^2} = \frac{2}{\pi} \frac{1}{2} 2 \int_{0}^{\infty} \frac{dy}{1+y^2} = \frac{2}{\pi} \mbox{ tan}^{-1}y ]_0^{\infty}
= 1.
\end{equation}

If we further generalize Eq.(1) by relaxing the constraint on $\beta$, we obtain the wave function
\begin{equation}
\psi(x,0) = A_o \mbox{ exp}\{[imx-c|x|^{\alpha}][1 + i\beta(z/|z|) \mbox{tan}(\pi \alpha /2)]\} .
\end{equation}
Note for the wave function in Eq.(21) that since $i$ multiplies $\beta$, the normalizing constant $A_o$ given in Eq.(2) is unchanged in terms of $\alpha$. For $\alpha = \frac{1}{2}$, which we will now consider, $A_o = c = (c')^{\frac{1}{2}}$.
Thus for $\alpha = \frac{1}{2}$ and $\beta = -1$, we obtain for the amplitude function the completely positive stable distribution (sometimes called Pearson V), multiplied by the normalizing constant $c$:
\begin{equation}
A(z) = A_o s_{\frac{1}{2},-1}(z;m,c) = c \frac{c}{\sqrt{(z-m)^3}} \mbox{ exp}[-\frac{c^2}{2(z-m)}] .
\end{equation}
As a check, we integrate the probability function $P(z) = [A(z)]^2$ corresponding to the amplitude function in Eq.(22):
\begin{eqnarray}
\int_{-\infty}^{\infty} [A(z]^2 dz = \int_{m}^{\infty} \frac{c^4}{(z-m)^3} \mbox{ exp}[-\frac{c^2}{(z-m)}] dz 
= \int_{0}^{\infty} \frac{1}{c^2} u^6 \frac{2c^2}{u^3} e^{-u^2} du\\
 = 2 \int_{0}^{\infty} u^3 e^{-u^2} = 2 \frac{1}{2} \Gamma(\frac{4}{2}) = 1,
\end{eqnarray}
where we have used the substitution $u = \frac{c}{(z-m)^{\frac{1}{2}}}$.

Finally, for the general case, we may express the amplitude function as a renormalized stable density, which is in turn
represented by a Taylor expansion in the form of gamma functions \cite{WF}[p. 583] (alternative expansions may be
found in \cite{HB}):
\begin{equation}
A(z) = A_o s_{\alpha,\beta}(z;0,1) = A_o \frac{1}{z} \sqrt{\frac{2}{\pi}} \sum_{k = 1}^{\infty} \frac{\Gamma(1+k/\alpha)}{k!}(-z)^k \mbox{ sin}[\frac{k\pi}{2\alpha}(\beta-\alpha)] ,
\end{equation}
for $z > 0$ and $1 < \alpha < 2$.  For $z < 0$ we have the general relation $s_{\alpha,\beta}(-z;m,c) =
s_{\alpha,-\beta}(z;m,c)$.  For $0<\alpha<1$ we have the similar expansion, for $z > 0$,
\begin{equation}
A(z) = A_o s_{\alpha,\beta}(z;0,1) = A_o \frac{1}{z} \sqrt{\frac{2}{\pi}} \sum_{k = 1}^{\infty} \frac{\Gamma(1+k\alpha)}{k!}(-z^{-\alpha})^k \mbox{ sin}[\frac{k\pi}{2}(\beta - \alpha)] .
\end{equation}
We may recover $m$ and $c$ in Eqs.(25,26) by the substitution $z = \frac{u-m}{c^{\frac{1}{\alpha}}}$.

\subsection{The uncertainty relation}

Now let's consider the uncertainty relation.  From Eq.(1), where the distribution is symmetric, we get the 
value for $(\triangle x)^2$ as:
\begin{equation}
(\triangle x)^2 = \int_{-\infty}^{\infty} \psi^* (x,0) x^2 \psi(x,0) dx .
\end{equation}
Inserting a factor of $u^2 = (2c)^{\frac{2}{\alpha}} x^2$ into the calculation of Eq.(5), we obtain
\begin{equation}
(\triangle x)^2 = \frac{1}{(2c)^{\frac{2}{\alpha}}} \frac{\Gamma(\frac{3}{\alpha})}{\Gamma(\frac{1}{\alpha})} .
\end{equation}
For $\alpha = 2$ this yields $(\triangle x)^2 = \frac{1}{4c}$, or in Heisenberg's formulation $\frac{1}{4\pi \tau}$ .

Next consider the uncertainty in $z$ (or $\sigma$).  First consider the case $\alpha = 2$.  From Eq.(15) we have
\begin{equation}
(\triangle \sigma)^2 = \int_{-\infty}^{\infty} (\sigma-\sigma_o)^2 [A(\sigma)]^2 d\sigma = (\frac{2}{\tau})^{\frac{1}{2}} \int_{-\infty}^{\infty} (\sigma-\sigma_o)^2 \mbox{ exp}[-\frac{2\pi(\sigma-\sigma_o)^2}{\tau}] d\sigma = \frac{\tau}{4\pi} .
\end{equation}
Thus we obtain the uncertainty relation
\begin{equation}
\triangle x \triangle \sigma = \frac{1}{4 \pi} .
\end{equation}
From the de Broglie relation $\triangle p = h \triangle \sigma$, where $h$ is Planck's constant, this becomes
\begin{equation}
\triangle x \triangle p = \frac{\hbar}{2} .
\end{equation}
However, for comparison with the results below, we will use for the (renormalized) Gaussian amplitude, the
uncertainty relation in the form
\begin{equation}
\triangle x \triangle z = \frac{1}{2} .
\end{equation}

Note that for the Cauchy density, where $\alpha = 1$, $\beta = 0$, the mean and variance don't exist ("are
infinite"). But we are considering a Cauchy \emph{amplitude}, and hence the \emph{square} of the Cauchy
density (renormalized) for the probability density.  For this density the second moment exists, as we will
now demonstrate.  From Eqs.(16,17), we calculate $(\triangle z)^2$ as:
\begin{eqnarray}
(\triangle z)^2 = \int_{-\infty}^{\infty} (z-m)^2 [A(z)]^2 dz = \frac{2}{\pi} \int_{-\infty}^{\infty} \frac{c^3 (z-m)^2}{[c^2 + (z-m)^2]^2} dz \\
= \frac{2c^2}{\pi} \int_{-\infty}^{\infty} \frac{y^2}{[1 + y^2]^2} dy = \frac{2c^2}{\pi} \frac{1}{2} 2 \int_{0}^{\infty} \frac{dy}{1+y^2} = c^2,
\end{eqnarray}
where we have used the relation
\begin{equation}
\int \frac{x^2 dx}{(ax^2+c)^n} = - \frac{1}{2(n-1)a} \frac{x}{(ax^2+c)^{n-1}} + \frac{1}{2(n-1)a} \int \frac{dx}{(ax^2+c)^{n-1}}.
\end{equation}
Thus we obtain the uncertainty relation, from Eqs.(28,34),
\begin{equation}
\triangle x \triangle z = \frac{1}{\sqrt{2}} .
\end{equation}

For the Pearson V amplitude, we have from Eqs. (22,23)
\begin{equation}
(\triangle z)^2 = \int_{-\infty}^{\infty} (z-m)^2 [A(z)]^2 dz = c^2 \int_{m}^{\infty} \frac{c^2}{(z-m)} e^{-\frac{c^2}{(z-m)}} dz = c^4 \int_0^{\infty} \frac{1}{y} e^{-y} dy ,
\end{equation}
where we have used the substitution $y = \frac{c^2}{(z-m)}$. This integral is divergent.  So instead we
calculate
\begin{equation}
\triangle z = \int_{-\infty}^{\infty} |z-m| [A(z)]^2 dz = \int_{m}^{\infty} \frac{c^4}{(z-m)^2} e^{-\frac{c^2}{(z-m)}} dz = c^2 \int_0^{\infty} e^{-y} dy = c^2.
\end{equation}
This yields, from Eqs.(28,38) the uncertainty relation
\begin{equation}
\triangle x \triangle z = \sqrt{\frac{15}{2}}.
\end{equation}
It is easy to see from Eq.(28) that the general uncertainty relation, as a function of $\alpha$, is 
\begin{equation}
\triangle x \triangle z = \sqrt{\frac{1}{(2)^{\frac{2}{\alpha}}} \frac{\Gamma(\frac{3}{\alpha})}{\Gamma(\frac{1}{\alpha})}}.
\end{equation}
This, then, is the reformulation of Heisenberg's uncertainty relation. The uncertainty is a function of the
characteristic exponent $\alpha$ of the (renormalized) stable amplitude. As $\alpha \rightarrow 0$, the
uncertainty becomes unbounded.

\subsection{The time-dependent wave function and the dispersion relation}

We can write the time-dependent wave equation corresponding to Eq.(1) as a superposition of plane waves:
\begin{equation}
\psi(x,t) = \int_{-\infty}^{\infty} A(z) \mbox{ exp}[i(zx-\nu(z) t)] dz,
\end{equation}
where $A(z)$ is the stable amplitude---a renormalized stable density, and $\nu(z)$ is the frequency.  A
dispersion relation connects $\nu(z)$ to $z$.

From the de Broglie relations
\begin{eqnarray}
E = h \nu \\
p = h z
\end{eqnarray}
we obtain the relation
\begin{equation}
\nu = z \frac{E}{p},
\end{equation}
which gives as the time-dependent wave equation
\begin{equation}
\psi(x,t) = \int_{-\infty}^{\infty} A(z) \mbox{ exp}[iz(x-\frac{E}{p} t)] dz.
\end{equation}
(Note that we do \emph{not} insert the classical relation $E = \frac{p^2}{2M}$, where $M$ is mass, at this point, because doing so does not yield a proper inverse Fourier transform.) Each plane wave equation $g(x,t) = \mbox{ exp}[iz(x-\frac{E}{p} t)]$ has differential operators $\frac{\partial^2 }{\partial x^2}$ and $\frac{\partial^2 }{\partial t^2}$ with eigenvalues $-z^2$ and $-z^2 \frac{E^2}{p^2}$ respectively:
\begin{eqnarray}
\frac{\partial^2 g}{\partial x^2} = -z^2 g \\
\frac{\partial^2 g}{\partial t^2} = -z^2 \frac{E^2}{p^2} g  .
\end{eqnarray}
These relations give rise to the partial differential equation
\begin{equation}
\frac{\partial^2 g}{\partial t^2} =  \frac{E^2}{p^2} \frac{\partial^2 g}{\partial x^2} .
\end{equation}

The time-dependent wave equation in Eq.(45) may be rewritten more fully (for $\alpha \ne 1$) as
\begin{equation}
\psi(x,t) = A_o \mbox{ exp}\{[im(x-\frac{E}{p} t)-c|(x-\frac{E}{p} t)|^{\alpha}][1 + i\beta((x-\frac{E}{p} t)/|(x-\frac{E}{p} t)|) \mbox{tan}(\pi \alpha /2)]\} .
\end{equation}
For symmetric distributions ($\beta = 0$), the probability density function corresponding to $\psi(x,t)$ is
\begin{equation}
P(x,t) = \psi^* (x,t) \psi(x,t) = A_o^2 \mbox{ exp}[-2 c|(x-\frac{E}{p} t)|^{\alpha}],
\end{equation}
which is the characteristic function of a stable density. For $\alpha = 2$, this is the Heisenberg time-dependent
density.

\subsection{Schr\"{o}dinger's equation revisted}

Schr\"{o}dinger's equation may be viewed as a simple consequence of the Heisenberg uncertainty relations.
Eq.(49) is a solution of the partial differential equation Eq.(48), so we have, as replacement for the
Schr\"{o}dinger equation, the partial differential equation
\begin{equation}
\frac{\partial^2 \psi}{\partial t^2} =  \frac{E^2}{p^2} \frac{\partial^2 \psi}{\partial x^2} ,
\end{equation}
which may be rewritten in the form
\begin{equation}
\nabla^2 \psi = \frac{1}{v^2} \frac{\partial^2 \psi}{\partial t^2}.
\end{equation}
This, of course, is the equation of a vibrating string, where $v = \frac{E}{p}$ is the speed of propagation of 
the waves.  It is a true wave equation, by contrast to Schr\"{o}dinger's heat equation formalism, which relates
$\frac{\partial \psi}{\partial t}$ to $\frac{\partial^2 \psi}{\partial x^2}$.
In fact, noting from Eq.(49), letting $\beta$ equal zero for simplicity, and letting sgn $y$ denote sgn $(x-\frac{E}{p} t)$, that
\begin{eqnarray}
\frac{\partial \psi}{\partial x} = (im-c\alpha|x-\frac{E}{p}t|^{\alpha-1} \mbox{ sgn }y) \psi \\
\frac{\partial^2 \psi}{\partial x^2} = ((im-c\alpha|x-\frac{E}{p}t|^{\alpha-1}\mbox{ sgn }y)^2-c\alpha(\alpha-1)|x-\frac{E}{p}t|^{\alpha-2})  \psi \\
\frac{\partial \psi}{\partial t} =  -\frac{E}{p} \frac{\partial \psi}{\partial x} \\
\frac{\partial^2 \psi}{\partial t^2} = \frac{E^2}{p^2} \frac{\partial^2 \psi}{\partial x^2} 
\end{eqnarray}
it does not appear to be particularly useful to relate $\frac{\partial \psi}{\partial t}$ to 
$\frac{\partial^2 \psi}{\partial x^2}$, although this can be done.  In fact,
\begin{equation}
\frac{\partial \psi}{\partial t} = - \frac{E}{p} \frac{(im-c\alpha|x-\frac{E}{p}t|^{\alpha-1}\mbox{ sgn }y)}{ ((im-c\alpha|x-\frac{E}{p}t|^{\alpha-1}\mbox{ sgn }y)^2-c\alpha(\alpha-1)|x-\frac{E}{p}t|^{\alpha-2})} \frac{\partial^2 \psi}{\partial x^2} .
\end{equation}
Only in the case of the Cauchy amplitude $\alpha = 1$ do we find this latter formulation in a simplified form:
\begin{equation}
\frac{\partial \psi}{\partial t} = - \frac{E}{p} \frac{1}{(im-c)} \frac{\partial^2 \psi}{\partial x^2} .
\end{equation}
If we now make the substitutions $E = \frac{p^2}{2M}$, $p = h \sigma$ we obtain
\begin{equation}
\frac{\partial \psi}{\partial t} =  \frac{h\sigma}{2M} \frac{im+c}{(m^2+c^2)} \frac{\partial^2 \psi}{\partial x^2} 
\end{equation}
which may be rewritten
\begin{equation}
ih \frac{\partial \psi}{\partial t} =  -\frac{h^2\sigma}{2M} \frac{m-ic}{(m^2+c^2)} \frac{\partial^2 \psi}{\partial x^2}. 
\end{equation}
It would appear that the traditional Schr\"{o}dinger equation involves a hidden Cauchy amplitude assumption. The
latter equation can be divided into two equations, one involving $m$ and the other involving $-ic$.

\subsection{Conclusion}

Stable distributions are the only distributions that exist as limit distributions of sums of random variables, thus giving rise to central limit theorems.  Therfore they play a paramount role in the physical world.  We have shown that
Heisenberg's original choice for a wave packet to illustrate his uncertainty principle is simply the
characteristic function (the inverse Fourier transform) of a Gaussian distribution, leading to a Gaussian amplitude function with $\alpha = 2$ and $\beta = 0$. Relaxing Heisenberg's assumptions to the general case $0<\alpha\le 2$, $|\beta|\le 1$, leads to stable amplitudes renormalized so that the integral of their squares are probability distributions. The renormalization constant gives rise to a new form of Heisenberg's uncertainty relation, expressed in terms of the characteristic exponent $\alpha$ of the underlying stable amplitude: $\triangle x \triangle z = \sqrt{\frac{1}{(2)^{\frac{2}{\alpha}}} \frac{\Gamma(\frac{3}{\alpha})}{\Gamma(\frac{1}{\alpha})}}$.  This relationship
was illustrated by explict calculation for the Gaussian ($\alpha =2$), the Cauchy ($\alpha = 1$), and the Pearson V
($\alpha = \frac{1}{2}$, $\beta = -1$).  As $\alpha \rightarrow 0$, the uncertainty $\triangle x \triangle z$ becomes
unbounded.  This means that, depending on the underlying stable amplitude, quantum uncertainty can arise at a macroscopic
level.

By eschewing the ad hoc classical insertion $E = \frac{p^2}{2M}$, we were able to solve for the time-dependent
wave equation as a superposition of plane waves, by taking the inverse Fourier transform of the stable amplitude function. For $\alpha = 2$, this recovers Heisenberg's case.  The wave function follows the partial differential equation
$\frac{\partial \psi}{\partial x^2} = \frac{1}{v^2} \frac{\partial^2 \psi}{\partial t^2}$, which is the equation 
for a vibrating string.  This is a proper wave equation, differing from Schr\"{o}dinger's equation, which is really a heat equation as it relates $\frac{\partial}{\partial t}$, instead of $\frac{\partial^2}{\partial t^2}$, to $\frac{\partial^2}{\partial x^2}$. The traditional form of the Schr\"{o}dinger equation can be recovered, but only in the case $\alpha = 1$. Thus it would appear that Schr\"{o}dinger's equation involves a hidden Caucy amplitude assumption.
This is not fatal, but is limiting.  The more general heat equation relationship is given by Eq.(57).

\end{document}